\documentclass[12pt,a4paper]{article}
\usepackage[utf8]{inputenc}
\usepackage[plainpages=false,pdfpagelabels=true]{hyperref}
\usepackage{amssymb,amsthm}
\usepackage[top=1cm,bottom=1cm,left=2 cm,right=2 cm]{geometry}
\usepackage{biblatex}
\usepackage{graphicx}
\usepackage{a4wide}  
\usepackage{amsfonts}
\usepackage{amsmath}
\usepackage{bbm}
\usepackage{booktabs} 
\usepackage{multicol}
\usepackage{ifpdf}

\ifpdf

\hypersetup{%
  pdftitle    = {Massive particle surfaces and black hole shadows from intrinsic curvature}
  pdfkeywords = {},
  pdfauthor   = {Oscar},
  plainpages  = true,
  colorlinks  = true,
  citecolor   = blue,
  urlcolor    = blue,
  linkcolor   = black
}

\makeatletter
\@addtoreset{equation}{section}
\makeatother

\begin{document}

\begin{center}
{\Large {\bf Massive particle surfaces and black hole shadows from intrinsic curvature}}

\vspace{1.5cm}

\renewcommand{\thefootnote}{\alph{footnote}}

\setcounter{footnote}{0}
\renewcommand{\thefootnote}{\arabic{footnote}}

\vspace{1cm}

\renewcommand{\thefootnote}{\alph{footnote}}
{\sl\large  Boris Berm\'{u}dez-C\'{a}rdenas \footnote{E-mail: bxbermudez[at]uc.cl }$^{b}$ and Oscar Lasso Andino}\footnote{E-mail: {oscar.lasso [at] udla.edu.ec}}
\setcounter{footnote}{0}
\renewcommand{\thefootnote}{\arabic{footnote}}

\vspace{1cm}
{\it $^a$ Facultad de Matem\'{a}ticas, Pontificia Universidad Católica de Chile,Avenida Vicu\~{n}a Mackenna, 4860, Santiago, Chile}
\end{center}

{\it $^b$ Escuela de Ciencias F\'{i}sicas y Matem\'{a}ticas, Universidad de Las Am\'{e}ricas, Redondel del ciclista, Antigua v\'{i}a a Nay\'{o}n, C.P. 170504, Quito, Ecuador}\\ \vspace{0.3cm}

\abstract{In a recent article PhysRevD.\textbf{111}.064001 (2025) a new geometric approach for studying massive particle surfaces was proposed. Using the Gaussian and geodesic curvatures defined over a two dimensional Riemannian metric, a criteria for the  existence of massive particle surfaces was provided. In this work, we generalize these results by including stationary spacetime metrics. We surmount the difficulty of having a Jacobi metric of the Randers-Finsler type by using a $2$-dimensional Riemannian metric that  is obtained by a dimensional reduction of the spacetime metric along the admitted Killing vectors. We provide a condition for the existence of massive particle surfaces and a simple characterization for null and timelike trajectories only by using intrinsic curvatures of a $2$-dimensional Riemannian surface. We study the massive particle surfaces of spacetimes that are not an asymptotically flat. We show that the Riemannian formalism can be used to study the shadows of the associated black holes. We show the existence of massive particle surfaces in the Kerr metric, the Kerr-(A)dS metric and in a solution, which is not asymptotically flat, of the Einsten-Maxwell-dilaton theory.}

\break

\section{Introduction}\label{sec1}
Black holes have attracted a lot of interest in the last few years. These objects have very interesting properties that have been studied theoretically and experimentally. After the experimental discovery of black holes in 2019 a lot of research has undergone \cite{EventHorizonTelescope:2019dse}. In particular, new techniques for characterizing and and describing the dynamics of these objects were developed \cite{Narayan:2019imo,Gralla:2019xty,Bronzwaer:2021lzo,Perlick:2021aok,Davelaar:2021eoi,Bogush:2022hop}. One of these new analytic approaches is a purely geometric one. Instead of studying the trajectories of particles by solving the geodesic equation, a more powerful method, based on intrinsic Riemannian curvatures, has been formalized in \cite{Qiao:2022jlu} where a criteria for the existence of null circular orbits, the so called light rings (LR's) was established. Using the Haddamard theorem with the optical metric\footnote{The optical metric is obtained by imposing the constraint $d\tau^2=-ds^2=0$, where $\tau$ is the proper time.} they defined a criteria for the trajectories that conform the photon sphere. In the trajectories where the geodesic curvature vanishes  the Gaussian curvature was calculated,  the sign of the Gaussian curvature  $\mathcal{K}\lessgtr 0$ helps to study the stability of the trajectory. These photon spheres were mathematically formalized in \cite{Claudel:2000yi}. The photon sphere is important for the dynamics of the photons, black hole shadows and gravitational waves \cite{Genzel:2024vou,Perlick:2021aok,Vertogradov:2024dpa,Ban:2024qsa,Chen:2024nua}. A generalization including a much bigger variety of metrics was done in \cite{Qiao:2022hfv}, see also  \cite{Cunha:2022nyw}.\\
We can think about the same approach but applied to timelike trajectories, in other words, for trajectories of massive particles, then a $2$-dimensional Riemannian metric that depends on the mass of the particle and the components of the spacetime metric is needed. An analogous concept to that of a photon sphere but for massive trajectories was developed in \cite{Bogush:2023ojz}, the so called massive particle surfaces MPS. Contrary to what happens with the photon spheres that are totally umbilic \cite{Senovilla:2011np,Okumura:1967}, the massive particle surfaces are partially umbilic \cite{Kobialko:2022uzj,Bogush:2024fqj}. However, from the point of view of a lower dimensional Riemannian metric, this partial umbilicity of the massive particle surfaces can be seen as total umbilicity \cite{Bermudez-Cardenas:2024bfi}. This Riemannian Jacobi metric was used to study the timelike and null geodesics of static black holes \cite{Gibbons:2015qja,Das:2016opi} and wormholes \cite{Arganaraz:2019fup,Duenas-Vidal:2022kcx}. \\
Our work uses a completely different approach for attacking the problem of determining the existence of massive particle surfaces. This new approach is powerful and simple and relies in the fact that we can build a Riemannian metric from a Lorentzian one by projecting over the directions of its Killing vectors. The results in \cite{Bermudez-Cardenas:2024bfi} cannot be directly generalized to stationary spacetimes, the principal problem is that the Jacobi metric that comes from a stationary spacetime is not Riemannian. Therefore, we cannot follow the geometric approach used for static spacetimes. The new results presented in this article go in that direction, namely, we follow a different way for finding a Riemannian metric and we show that we can use that Riemannian metric to determine the existence of a  (MPS). Moreover, we show that  the information regarding the geodesics and shadows of the original Lorentzian metric are inherited to  this new $2-$dimensional Riemannian metric. Moreover, our results can be applied to non-asymptotically flat spacetimes.
The results in \cite{Bermudez-Cardenas:2024bfi} are restricted to spacetimes whose metrics are defined by $g_{tt}=f(r),g_{rr}=1/g(r),g_{\theta\theta}=r^2$ and $g_{\phi\phi}=r^2\sin^2(\theta)$ and such that at spatial infinity the spacetime metric becomes flat. In this article we present a generalization,  for stationary spacetime metrics, of the formalism developed in \cite{Bermudez-Cardenas:2024bfi}. We study asymptotically AdS/dS spacetimes and analyze the existence of circular null/timelike geodesics. Our conclusions are also a generalization of the null case presented in \cite{Qiao:2022jlu,Qiao:2022hfv}. First, we develop our technique for  spacetime metrics that are not necessarily asymptotically flat and without any restriction over the components of the metric. Second, we generalize to spacetime metrics that are stationary. In this case we have the inconvenience that  the Jacobi metric is of the Randers-Finsler type \cite{Caponio:2009es,Li:2021xhy,Gibbons:2008zi,Werner:2012rc,Ono:2017pie}. In order to avoid this type of geometry we use the approach developed in \cite{Arganaraz:2021fwu} for building a $2$-dimensional Riemannian metric. This new metric is obtained by the projection of spacetime metric over surfaces of constant energy $E$ and constant momentum $L$ . Using this $2$-dimensional metric we can calculate the geodesic and Gaussian curvatures. When the geodesic curvature vanishes we obtain a condition for the existence of massive particle surfaces: the master equation proposed in \cite{Bogush:2023ojz}. We also get more, we are able to characterize the horizon of the black hole using Riemannian geometry. Moreover, we calculate the radius of the photon sphere and the radius of the innermost stable circular orbit (ISCO). We also calculate  the parameters that characterize the shadows of the black hole showing that the information regarding the black hole shadows can be also deduced from curvatures of the Jacobi metric. \\

In section \ref{s:1} we present a brief review of massive particle surface formalism and discuss the difference with photon spheres. In section \ref{s:2}  we present a brief review of Gaussian and geodesic curvatures in the context of $2$-dimensional Riemannian surfaces. In subsection \ref{s:3}, we show how the Jacobi metric and its curvatures can be calculated for general metrics. In section \ref{s:4} we study the massive particle surfaces for stationary spacetimes. In section \ref{s:5} we particularize for three types of metrics, namely the Kerr metric, the Kerr-(A)dS and a solution of Einstein-Maxwell-dilaton (EMD) theory.  

\section{Massive particle surfaces and photon spheres}\label{s:1}
Massive particle surfaces (MPS) are built with timelike trajectories and have been mathematically formalized recently \cite{Bogush:2023ojz,Kobialko:2022uzj}. MPS are natural generalizations of photons surfaces which are built with null trajectories and formalized in \cite{Claudel:2000yi}.  More specifically, any worldline of a massive particle  with total energy $E$, with mass $m$ and electric charge $q$ which is initially tangent to $S_{MPS}$ will remain tangent to it. When $m=0$ and $q=0$ the MPS reduces to a photon sphere $S_{PS}$, however now we require that the total energy $E$ become arbitrary. In the photon sphere $S_{PS}$ any null geodesic initially tangent to $S_{PS}$ will remain tangent to it, this leads to the fact that $S_{PS}$ is an umbilic surface, technically this means that the first fundamental form is proportional to the second fundamental form of the manifold. On the other hand, the massive particle surface $S_{MPS}$ has the property of being partially umbilic\footnote{The photon surface of the Kerr spacetime is also partially umbilic \cite{Kobialko:2020vqf}.}. The photon surfaces are very important  when studying the gravitational shadows of the black holes. The timelike trajectories and the geometry of the surfaces that they form are important when accretion disks around black holes or other compact objects are studied. Due to the strong gravitational fields around massive objects the probability of forming an accretion disk is high. A very simple example is the thin accretion disk formed around a Schwarzschild black hole \cite{Narayan:2019imo}. The image of a black hole is altered when massive particles surrounding the black hole form an accretion disk \cite{Gralla:2019xty}, where an influence over the shadow of the black hole has been detected \cite{Gralla:2019xty,Bronzwaer:2021lzo,Olmo:2023lil,Peng:2020wun}. A general model of a thin accretion disk can be built by considering a free, electrically neutral plasma orbiting along a timelike geodesic in the equatorial plane $\theta=\pi/2$. Using the radius of the innermost stable circular orbit $r_{ISCO}$ the accretion disc can be divided into two zones, the inner and the outer zone, where the motion of massive particles will be different in each zone \cite{Li:2024ctu, Zheng:2024ftk} . Thus, the accretion disk flows move following timelike circular orbits, the orbits that conform a massive particle surface. Note that our formalism will provide a geometric origin of the usual calculation that involves the use of the effective potential of the radial geodesic equation. Take as an example  $r_{isco}$, in the next section we will see how it can be derived from the master equation of a MPS. The condition for the existence of circular timelike orbits is also derived from the geodesic curvature. The calculations are easier compared with those involving the geodesic equation, and we do not need to request integrability of the geodesic equation.
In this section we introduce the mathematical definition of massive particle surfaces and the criteria for determining its existence.\\
Let us start by considering a $n-$dimensional spacetime whose metric is $g_{\mu\nu}$ and such that it has a timelike Killing vector $k^{\mu}$. The worldline of a particle is noted with $\gamma^{\mu}$, then it has to satisfy
\begin{equation}\label{econd}
\dot{\gamma}^{\mu}\nabla_{\mu}\dot{\gamma}^{\nu}=0,\,\,\,\,\dot{\gamma}^{\mu}\dot{\gamma}_{\mu}=-m^2,\,\,\,\,\mathcal{E}_{k}=-k_{\mu}\dot{\gamma}^{\mu},
\end{equation}
where $\mathcal{E}_k$ is the total energy\footnote{When an electromagnetic field is considered by introducing an electromagnetic potential $A_{\mu}$ the total energy $\mathcal{E}$ has two contributions, the kinetic energy defined in \eqref{econd} and the potential energy $\mathcal{E}_p=-qk_{\mu}A^{\mu}$. }. \\
A massive particle surface is a $(n-1)$-dimensional timelike surface $\mathcal{S_{M}}$, such that any tangent vector to $\mathcal{S_{M}}$ remains tangent to $\mathcal{S_{M}}$. In other words, a particle located at any point in $S_{M}$ such that its velocity is tangent to $S_{M}$ then it remains tangent forever. Thus, for a spacelike outer normal vector $n_{\mu}$  we have:
\begin{equation}
h_{\mu\nu}=g_{\mu\nu}-n_{\mu}n_{\nu},\,\,\,\,\chi_{\mu\nu}=h_{\mu}^{\alpha}h_{\nu}^{\beta}\nabla_{\alpha}n_{\beta}.
\end{equation}
The projection of the timelike Killing vector $k^{\mu}$ onto the surface $\mathcal{S_{M}}$ is given by $\kappa^{\mu}=h_{\alpha}^{\mu}k^{\alpha}$. In the static case $\kappa^{\mu}=k^{\mu}$ then if $k_{\mu}k^{\mu}\neq 0$ then \cite{Kobialko:2022uzj}
\begin{equation}\label{ineq:1}
0<|k^{\mu}k_{\mu}|\leq \frac{\mathcal{E}^2}{m^2}
\end{equation}

The inequality \eqref{ineq:1} reveals a restriction over the motion on  massive particle surfaces. This restriction is not present when a photon surface is considered. Thereby a study of timelike geodesics can be done only by studying the geometry of the massive particle surfaces,  without solving the geodesic equations. The relation between the induced metric over the surface an the extrinsic curvature is given by
\begin{equation}\label{puc}
\chi_{\mu\nu}=\frac{\chi_{\tau}}{n-2}H_{\mu\nu},\,\,\,\,\,H_{\mu\nu}=h_{\mu\nu}+\frac{m^2}{\mathcal{E}^2}\kappa_{\mu}\kappa_{\nu}.
\end{equation}
The conditions for the existence of MPS can be written in a relative simple way using the so called master equation \cite{Kobialko:2022uzj}
\begin{equation}\label{master:1}
\mathcal{E}_{\pm}=\pm \sqrt{\frac{\kappa^2 \chi_{\tau}}{W}},
\end{equation}
where
\begin{equation}
W=-\chi_{\tau}+\frac{n-2}{2}n^{\mu}\nabla_{\mu}\ln \kappa^2,\,\,\,\,\, \chi_{\tau}=\frac{n-2}{H}\chi_{\mu}^{\mu}.\\
\end{equation}
If an electromagnetic field is added the formalism has to be modified, although the definitions remain. In order to show that the surface $\mathcal{S_M}$ is  a MPS the right side of the equation \eqref{master:1} has to be constant. Moreover, the condition $d\mathcal{E}/dr=0$ leads to a equation that is solved by the  radius of the innermost stable circular orbit (ISCO).\\
The geometric definition of a MPS leads to the so called partial umbilicity property. Thus, if we denote the system of vectors which are tangent to $S_M$ but orthogonal to  $k^{\mu}$ as $\tau_{(j)}^{\mu}$, with $j=1,2,...,n-2.$, then equations \eqref{puc} lead to the condition

\begin{eqnarray}
\tau_{(j)}^{\alpha}\tau_{(j)}^{\beta}&=&\frac{\chi_\tau}{n-2}\tau_{(j)}^{\mu}\tau_{(j)}^{\nu}h_{\mu\nu}\label{umbi:1}\\
0&=&n^{\mu}\tau^{\nu}_{(j)}\left(\frac{\nabla_\nu k_{\mu}}{k^\mu_{\mu}}\right)\label{umbi:2}.
\end{eqnarray}
Both equations \eqref{umbi:1} and \eqref{umbi:2} reveal two facts that are going to be important for our study. The first one is related with the partial umbilicity of the surface $S_{M}$. The property that the first and second fundamental forms of the metric are proportional (not equal) is called partial umbilicity. In the static case, the photon surface is called partially umbilic, However, when the metric is projected over the directions of its Killing vectors, the resulting Riemannian metric is totally umbilic, namely the first and the second fundamental form are the same. Due to the fact that this equality is valid only for their values on tangent vectors that are orthogonal to the Killing vector, we can project over the directions of its Killing vectors, then partial umbilicity becomes total umbilicity. As the reader might be aware of, the calculations can be really complicated. Moreover, the geodesic equation can be  very difficult to integrate, therefore new geometric methods are needed. Using only the Jacobi metric and its curvature we will be able to determine the existence of MPS, determine the $r_{ISCO}$, the photon sphere radius $r_{ph}$ and to classify the trajectories of null and timelike particles without resting in the geodesic equations.\\
If we project the spacetime metric over surfaces of constant energy, the resulting metric is Riemannian, and from the point of view of this metric the  partial umbilicity property is seen as a total umbilicity  \cite{Bermudez-Cardenas:2024bfi}. The Riemannian metric is called the Jacobi metric and it has very interesting properties that can be used using the powerful results developed in Riemannian geometry. In the following sections we  show how to calculate the geodesic and Gaussian curvatures using the Jacobi metric.

\section{Gaussian and geodesic curvature on circular orbits}\label{s:2}
A $2$-dimensional surface can have different curvatures. We are interested in curvatures that are called intrinsic: those that are independent of the ambient space where the surface is embedded. The intrinsic curvatures are independent of the coordinate system and therefore we can choose a type of  coordinates conveniently. We consider a $2$-dimensional $S$ surface on which we have defined two curving linear coordinates $x_1$ and $x_2$, then the length of a curve defined over $S$ can be calculated using a Riemannian metric  $G_{\mu\nu}$:
\begin{equation}\label{2dm}
ds^2=G_{11}dx_{1}^2+G_{22}dx_2^2.
\end{equation}

Thus, for any curve $\alpha(t)=(x_1(\lambda),x_2(\lambda))$ living on the surface $S$, where $\lambda$ is the arclength parameter, its geodesic curvature can be written \cite{Berg:1988,DCarmo:2000}
\begin{equation}\label{geogen:1}
\kappa_{g}(\alpha(\lambda))=\frac{d\sigma}{ds}-\frac{1}{2\sqrt{g_{22}}}\frac{\partial \log(g_{11})}{\partial x_2}\cos(\sigma)
+\frac{1}{2\sqrt{g_{11}}}\frac{\partial \log(g_{22})}{\partial x_1}\sin(\sigma),
\end{equation}
where $\sigma$ is the angle between the tangent vector $\dot{\alpha}$ and the axis $x_1$. The geodesic curvature \eqref{geogen:1} measures how far is the curve $\alpha(\lambda)$ from being a geodesic. In other words, when $\kappa_g(\alpha(\lambda))=0$ the curve $\alpha(\lambda)$ is a geodesic in the surface $S$. Thus, in the following we will use the term geodesic curvature of a metric, meaning the geodesic curvature of an arbitrary curve $\alpha(\lambda)$ defined over the surface\footnote{Mathematically, it can be shown that \cite{Qiao:2022jlu}
 \begin{equation}
 \kappa_g(\alpha (\lambda))=0 \Leftrightarrow \frac{d^2x_{i}}{d\lambda^2}-\Gamma^{i}_{jk}\frac{dx_j}{d\lambda}\frac{dx_k}{d\lambda}\bigg\vert_{\alpha=\alpha(\lambda)}=0.
 \end{equation}} $S$. For a circular orbit we have $\sigma=\pi/2$, therefore the geodesic curvature of a circular orbit becomes
 \begin{equation}
 \kappa_{g}(\alpha(\lambda))\bigg\vert_{r=r_{c}}=\frac{1}{2\sqrt{g_{11}}}\frac{\partial \log(g_{22})}{\partial x_1}\bigg\vert_{r=r_{c}},
 \end{equation}
 where $r_c$ stands for the radius of a circular orbit. The condition 
\begin{equation}
\kappa_{g}(\alpha(\lambda))\bigg\vert_{r=r_{c}}=0,
\end{equation}
characterizes the existence of circular geodesics, in the null case they are the very well known LR's. The only input needed is a Riemannian metric defined over the $2$-dimensional surface $S$.\\
The Gaussian curvature can be calculated using the Brioschi prescription. Thus, when the metric is given by \eqref{2dm} the Brioschi formula reads \cite{Berg:1988,DCarmo:2000}
\begin{equation}\label{gaussgen:1}
\mathcal{K}= -\frac{1}{2\sqrt{G_{11}G_{22}}}\left(\frac{\partial}{\partial x_2}\left(\frac{\partial_{x_{2}}G_{11}}{\sqrt{G_{11}G_{22}}}\right)
\right. 
 \left. +\frac{\partial}{\partial x_1}\left(\frac{\partial_{x_{1}}G_{22}}{\sqrt{G_{11}G_{22}}}\right)\right)
\end{equation}

The Gaussian curvature \eqref{gaussgen:1} is an intrinsic property of the surface $S$. This curvature measures how far a surface is of being flat. As in the case of geodesic metric, the only input needed is a Riemannian metric defined in $S$. It is important to note that Gaussian curvature \eqref{gaussgen:1} can be calculated for any metric of the type \eqref{2dm} without any assumption about the type of orbits. We are going to use the geodesic and Gaussian curvatures to study the existence of light rings and timelike circular orbits in stationary black holes.

\section{The Riemannian curvature of spacetime}\label{s:3}
We summarize the results about the Gaussian and geodesic curvatures of the Jacobi metric, a Riemannian metric obtained by projecting the spacetime metric over surfaces of constant energy and constant momentum. We are going to use the Riemannian geometric approach  proposed in \cite{Arganaraz:2021fwu} and the general results for the curvatures of the Jacobi metric presented in \cite{Bermudez-Cardenas:2024bfi}, where  the particular case $g_{tt}=f(r),g_{rr}=1/g(r), g_{\theta\theta}=h(r)$ and $g_{\phi\phi}=h(r)\sin^{2}(\theta)$ was analyzed.  We are going to generalize  the results of \cite{Bermudez-Cardenas:2024bfi} to  metrics of the type:

\begin{equation}\label{staticm:1}
ds^2=g_{tt}(r) dt^2+g_{rr}(r)dr^2+g_{\theta\theta}(r)d\theta^2+g_{\phi\phi}(r,\theta)d\phi^2.
\end{equation}

The Jacobi metric obtained by the dimensional reduction of \eqref{staticm:1} along the Killing vector\footnote{A more conceptual way of seeing it is that the projection is made over surfaces of constant energy.} $\partial_t$   is given by
\begin{equation}\label{jmstatic:1}
J_{ij}dx^{i}dx^{j}=\left(\delta-\frac{\mathcal{E}^2}{g_{tt}(r)}\right)g_{ij},
\end{equation}
where $g_{ij}$ is a a $3$-dimensional Riemannian metric\footnote{If we use spherical coordinates $i,j=r,\theta,\phi$}. The metric \eqref{jmstatic:1}, when restricted to the equatorial plane, becomes a two dimensional metric and therefore is defined over a $2$- dimensional surface $S$. When  the metric \eqref{jmstatic:1} is conformally flat, which is the case when spherically symmetric spacetimes are considered, the Gaussian curvature $\mathcal{K}$ can be easily calculated \cite{Gibbons:2015qja,Bermudez-Cardenas:2024bfi}. The surface $S$ has intrinsic curvature that is measured by the Gaussian curvature, which measures how far $S$ is from being flat. In general, it is difficult to determine the sign of $\mathcal{K}$. However, there are particular cases, such as null circular orbits, the so called light rings, where it can be done \cite{Cunha:2022nyw,Qiao:2022jlu,Qiao:2022hfv}, then the curvature $\mathcal{K}$ can be used to study the stability of a trajectory. Similarly, the geodesic curvature  measures how far, a curve defined over $S$, is of being a geodesic. If the geodesic curvature vanishes over a curve, then that curve is a geodesic. Certainly, if the surface $S$ is a sphere, then the vanishing of the geodesic curvature leads to the existence of circular orbits.\\
The Jacobi metric \eqref{jmstatic:1} can be rewritten

\begin{equation}\label{jacobig:1}
J_{ij}dx^{i}dx^{j}=\left(\frac{E^2-m^2g_{tt}(r)}{g_{tt}(r)}\right)\left(g_{rr}(r)dr^2+g_{\theta\theta}(r)d\theta^2+g_{\phi\phi}(r,\theta)d\phi^2\right),
\end{equation}

where $i,j=r,\theta,\phi$. Because of spherical symmetry we can assume  $\theta=\pi/2$, thereby obtaining a two dimensional Riemannian metric, whose Gaussian curvature is given by

\begin{eqnarray}\label{gaussianc:2}
\mathcal{K}&=&\sqrt{\frac{\mathfrak{g}_{\phi\phi}}{g_{rr}}}\left(\frac{F''}{F^2}+\frac{2F'}{F}\frac{\mathfrak{g}'_{\phi\phi}}{\mathfrak{g}_{\phi\phi}}+\frac{\mathfrak{g}''_{\phi\phi}}{\mathfrak{g}_{\phi\phi}}+\frac{(F')^{2}}{F^2}+\frac{1}{F}\frac{\mathfrak{g}'_{\phi\phi}}{\mathfrak{g}_{\phi\phi}}
\right.\nonumber\\
& &
\left.
+\left(\frac{F'}{F}+\frac{\mathfrak{g}'_{\phi\phi}}{\mathfrak{g}_{\phi\phi}}\right)(g_{rr}\mathfrak{g}_{\phi\phi})'\right).
\end{eqnarray}

where $\mathfrak{g}_{\phi\phi}=\left. g_{\phi\phi}\right|_{\theta=\pi/2}$. It is clear that when $g_{rr}=1/g(r), g_{\phi\phi}=r^2\sin(\theta)$ we recover equation $(3.15)$ in \cite{Bermudez-Cardenas:2024bfi}. Moreover, if we also set $h(r)=r^2$ and $m=0$ we obtain the results in \cite{Qiao:2022hfv}. Using \eqref{gaussianc:2} together with the criteria developed in \cite{Qiao:2022hfv} for classifying  the stability of  the photon orbits a study of the stability of the trajectories of massive particles can be carried out. \\
The geodesic curvature  of the metric \eqref{jmstatic:1} is given by

\begin{equation}\label{geo:1}
\kappa_{g}=\frac{1}{2F^{3/2}g_{rr}^{1/2}g_{\theta\theta}^{1/2}}\frac{\partial \left(F \mathfrak{g}_{\phi\phi}\right)}{\partial r},
\end{equation}
The geodesics are going to be the curves $\alpha(t)$ such that $\kappa_g(\alpha(t))=0$ and therefore, from \eqref{geo:1} we get
\begin{equation}
-\partial_r g^{tt} \mathfrak{g}_{\phi\phi}-g^{tt}\partial_r \mathfrak{g}_{\phi\phi}-\frac{m^2}{E^2}\partial_{r}\mathfrak{g}_{\phi\phi}=0,
\end{equation}
 hence
\begin{equation}\label{cond:2}
\frac{E^2}{m^2}=\frac{g_{tt}^2 \partial_ r \mathfrak{g}_{\phi\phi}}{g_{tt}\partial_r \mathfrak{g}_{\phi\phi}-\partial_r g_{tt}}
\end{equation}
The condition \eqref{cond:2} leads to a criteria for the existence of massive particle surfaces, if the right hand side of \eqref{cond:2} is constant then there is a massive particle surface. Thus, the condition for the existence of circular orbits of the metrics of the type \eqref{staticm:1} is given by

\begin{equation}\label{mpsc:1}
\mathfrak{g}_{\phi\phi}\partial_r F+ F \partial_r \mathfrak{g}_{\phi\phi}=0,
\end{equation} 
and again, equation \eqref{mpsc:1} reduces to equation $(5.1)$ in \cite{Bermudez-Cardenas:2024bfi} when $g_{\phi\phi}=h(r)$ and $g_{rr}=1/g(r)$. The condition \eqref{mpsc:1} can be applied to any metric of the form \eqref{staticm:1}. In particular, for a static spherically symmetric asymptotically flat spacetime  metric, the procedure described in \cite{Bermudez-Cardenas:2024bfi} leads to a metric that is defined over the tangent space spanned by the vectors $\partial_r$ and $\partial_\theta$.  The condition \eqref{mpsc:1} together with the sign of the Gaussian curvature \eqref{gaussianc:2} can be used to study the stability of the geodesics followed by massive particles. These findings represent a generalization of the criteria presented in \cite{Qiao:2022jlu}, where all results where stated for null trajectories. \\
As we have shown, a generalization to stationary spacetimes can be done, in the next section we are going to show how  results presented in this section can be generalize to  these stationary spacetimes. 
The condition \eqref{mpsc:1} leads to a simple and powerful  method for calculating what is known as the master equation for the existence of a massive particle surfaces.\cite{Bogush:2023ojz,Kobialko:2022uzj,Bermudez-Cardenas:2024bfi}.

\subsection{A Riemannian metric from stationary spacetimes}\label{s:4}

The generalization described in the previous section can be extended even more when stationary spacetimes are considered. A formal approach can be found in \cite{Kobialko:2020vqf}. Nevertheless, the optical metric obtained  is of the Randers-Finsler type \cite{Werner:2012rc,Ono:2017pie} and the Jacobi metric for massive particle trajectories is of the same type \cite{Gibbons:2008zi,Li:2021xhy,Caponio:2009es}. If we want to extend the results of previous sections to a stationary metric we need a Riemannian metric. There is a procedure that brings a Riemannian metric from \eqref{staticm:1} by dimensional reduction along the Killing vectors of the spacetime metric. The resulting metric inherits the information needed and in the static case it reduces to known results. In this section we present the formalism developed that helps to obtain a Riemannian metric from a Lorentzian metric. We present a review of the procedure detailed in \cite{Arganaraz:2021fwu}.\\
It is known that the geodesics defined in a spacetime with metric $g_{\mu\nu}$ can be deduced form the Lagrangian
\begin{equation}
\mathcal{L}=\frac{1}{2}g_{\mu\nu}\dot{x}^{\mu}\dot{x}^\nu,
\end{equation}
then the Hamiltonian of the theory is written as $\mathcal{H}=\Sigma_{\mu}p_{\mu}\dot{x}^{\mu}-\mathcal{L}$ where $p_{\mu}=\frac{\partial \mathcal{L}}{\partial \dot{x}^{\mu}}=\dot{x}_{\mu}$. On the other side, the corresponding Hamilton-Jacobi equation can be written 
\begin{equation}\label{hj:1}
\frac{\partial S}{\partial \lambda}+\frac{1}{2}g^{\mu\nu}p_{\mu}p_{\nu}=0,
\end{equation}
where $p_{\mu}=\partial S/\partial x^{\mu}$ and\footnote{The action $S$ and the Lagrangian $\mathcal{L}$ satisfy the identity $\frac{\partial S}{\partial x^{\mu}}=\frac{\partial \mathcal{L}}{\partial \dot{x}^{\mu}}$} $S=\int \mathcal{L}dt$. Moreover, a geodesic tangent vector $\dot{x}^{\mu}$ has a norm $\delta$ which satisfies
\begin{equation}
\delta=g_{\mu\nu}\dot{x}^{\mu}\dot{x}^{\nu}=2\mathcal{L}=2\mathcal{H},
\end{equation}
where $\delta=0$ for null geodesic and $\delta=-m^2$ for timelike geodesics. From equation \eqref{hj:1} and the previous expression we obtain
\begin{equation}\label{hj:2}
-\frac{\delta}{2}+\frac{1}{2}g^{\mu\nu}p_{\mu}p_{\nu}=0.
\end{equation}
We consider a $4$-dimensional spacetime such that $x^{\mu}=(t,x^{i}),\,\,i=1,2,3$, then equation \eqref{hj:2} leads to
\begin{equation}
-\frac{\delta}{2}+\frac{1}{2}(g^{tt}p_{t}p_{t}+g^{ij}p_{i}p_{j})=0,
\end{equation}
hence
\begin{equation}
\left(\frac{1}{\delta-g^{tt}p_tp_t}\right)g^{ij}p_{i}p_j=1.
\end{equation}
We take the previous condition as a definition of the Jacobi metric, namely $J^{ij}p_{i}p_{j}=1$ where
\begin{equation}
J^{ij}=\left(\frac{1}{\delta-g^{tt}p_tp_t}\right)g^{ij},
\end{equation}
which has an inverse
\begin{equation}\label{gjm:3}
J_{ij}=\left(\delta-g^{tt}p_{t}p_{t}\right)g_{ij}.
\end{equation}
The metric $J_{ij}$ is a Riemannian metric and was first proposed in \cite{Arganaraz:2021fwu}. This metric is a three dimensional Riemannian metric and carries some information regarding the geodesics of the metric $g_{\mu\nu}$. The procedure can be continued by dimensionally reducing along any Killing vector. In general, we can take the indices that run over the coordinates having a Killing vector as $\alpha,\beta,\gamma,..$.  The coordinates on which there is not Killing vectors are named $a,b,c...$, then the Jacobi metric reads
\begin{equation}\label{gjm:4}
J_{ab}=\left(\delta-g^{\alpha\beta}p_{\alpha}p_{\beta}\right)g_{ab}.
\end{equation}
The previous equation shows that if the dimensional reduction is made along a  Killing vector then $p_{\alpha}$ is a constant, and therefore the  conformal factor of the metric \eqref{gjm:4} is independent of time. The metric  \eqref{gjm:4} is a Riemannian metric and therefore we can use its geometric properties, such as the geodesic or Gaussian curvatures, for studying massive particle surfaces, photons spheres and the shadows. In the following, we will use this procedure to find a Riemannian metric for studying the massive particle surfaces for general stationary spacetimes.

\subsection{Partial umbilicity condition in stationary spacetimes}

The most general metric for a stationary spacetime with two isometries $\partial_t, \partial_\phi$, in $4$-dimensions is written:

\begin{equation}\label{staticm:2}
ds^2=g_{tt}(r,\theta) dt^2+g_{rr}(r,\theta)dr^2+2g_{t\phi}(r,\theta)dt d\phi
+g_{\theta\theta}(r,\theta)d\theta^2+g_{\phi\phi}(r,\theta)d\phi^2.
\end{equation}

A Riemannian geometric approach for studying geodesics was developed in \cite{Arganaraz:2021fwu}, where for a metric of the type \eqref{staticm:2} a Riemannian metric was presented:

\begin{equation}\label{jacobim:3}
J_{ij}dx^{i}dx^{j}=F(r,\theta)g_{ab},
\end{equation}
where $a,b=r,\theta$ and 
\begin{equation}
F(r,\theta)=\delta-\frac{1}{-g_{t\phi}^2+g_{\phi\phi}g_{tt}}\left(g_{\phi\phi}E^2+2g_{t\phi} EL+g_{tt}L^2\right)
\end{equation}  
The Gaussian curvature of the metric \eqref{jacobim:3} is given by

\begin{equation}\label{gaussiank:1}
\mathcal{K}=\frac{Z(r,\theta)}{4 F^3 g_{\theta \theta
   }^2 g_{rr}^2}
\end{equation}

\begin{eqnarray}\label{gaussiank:2}
Z(r,\theta)&=&Fg_{\theta \theta } \left(g_{\theta \theta } \left(\partial_{\theta}F \partial_{\theta}g_{rr}+\partial_r F
   \partial_r g_{rr}\right)+F \left(\partial_{\theta}g_{\theta \theta }\partial_{\theta}g_{rr}+\partial_r g_{\theta \theta }
   \partial_r g_{rr}\right)\right)\nonumber\\ 
  & & -F g_{rr}g_{\theta \theta } \left(\partial_{\theta}F \partial_{\theta}g_{\theta \theta }+\partial_{r}F \partial_{r}g_{\theta \theta}+2 F \left(\partial_{\theta\theta}g_{\theta \theta }+\partial_{rr}g_{\theta \theta }\right)\right)\nonumber\\
& &+2 g_{rr}g^2_{\theta \theta }\left((\partial_{\theta}F)^2+(\partial_r F)^2-F \left(\partial_{\theta\theta}F+\partial_{rr}F\right)\right) \nonumber\\
& &+g_{rr}F^2 \left((\partial_{\theta}g_{\theta \theta
   })^2+(\partial_{r}g_{\theta \theta })^2\right).
 \end{eqnarray}

On the other side, the geodesic curvature calculated with \eqref{jacobim:3} is given by
\begin{equation}\label{geode}
\kappa_{g}=\frac{g_{\theta\theta}\partial_r F(r,\theta)+F(r,\theta)\partial_{r} g_{\theta\theta}}{2 F^{3/2}(r,\theta)g_{rr}^{1/2}g_{\theta\theta}}.
\end{equation} 

For a circular geodesic we have that $\kappa_{g}=0$, then from the previous equation we obtain

\begin{equation}\label{cond:4}
\partial_r F(r,\theta)g_{\theta\theta}+F(r,\theta)\partial_r g_{\theta\theta}=0.
\end{equation}

The condition \eqref{cond:4} represents the generalization, for stationary spacetimes, of the partial umbilicity condition found in \cite{Bermudez-Cardenas:2024bfi}, and therefore it encodes the information about the massive particle surface. \\
Let us analyze a particular aspect of the metric \eqref{jacobim:3}, namely the conformal factor $F(r,\theta)$ dependence on angular coordinates. In the static, spherically symmetric asymptotically flat spacetimes the metric \eqref{staticm:2} becomes:

\begin{equation}\label{staticm:4}
ds^2=f(r)dt^2+\frac{dr^2}{f(r)}+r^2d\theta^2+r^2\sin(\theta)d\phi^2,
\end{equation}
 and the Jacobi metric \eqref{jacobim:3} transforms to
\begin{equation}
J_{ij}dx^{i}dx^{j}=F(r,\theta)\left(\frac{dr^2}{f(r)}+r^2 d\theta^2\right),
\end{equation}

where
\begin{equation}
F(r,\theta)=\left(\delta+\frac{E^2}{f(r)}-\frac{L^2}{r^2\sin^2(\theta)}\right).
\end{equation}
The conformal factor  $F(r,\theta)$ depends on both coordinates\footnote{The conformal factor in the static case is parametrized by $r,\phi$. } $r,\theta$. The condition \eqref{cond:4} becomes:\\

\begin{equation}\label{staticmsp:1}
\frac{E^2}{m^2}=\frac{2f^2}{2f-f'r}.
\end{equation}

It is evident that the dependence of the Jacobi metric on the coordinate $\theta$ is not an inconvenient in the static case. The calculation is simple and direct and a massive particle surface can be defined. 
We have shown that the Riemannian metric \eqref{jacobim:3} carries the same information regarding the massive particle surfaces as the Jacobi metric defined over projection on surfaces of constant energy in the equatorial plane $\theta=\pi/2$.

The condition \eqref{cond:4} depends on the component $g_{\theta\theta}$ of the metric  and the Riemannian surface is parametrized by the coordinates $r,\theta$. The equation \eqref{cond:4} is a generalization of the condition obtained in \cite{Bermudez-Cardenas:2024bfi}.

\section{Examples}\label{s:5}
In this section we provide some examples of spacetime metrics that are stationary and in certain cases non asymptotically flat. The master equation for the existence of massive particle surfaces is calculated using the Riemannian metric defined in \eqref{jacobim:3}. We also show that this metric inherits the information regarding the shadows of black holes, namely the  parameters that can be used to construct the so called celestial coordinates  that are used to describe the shape of the shadow. We also calculate, when it is possible, the radius of the innermost stable circular orbit and the photon surface radius.

\subsection{Kerr spacetime}  
The Kerr metric is the prototypical example of  a stationary spacetime. Its null and timelike geodesics have been widely studied \cite{Teo:2003ltt,Teo:2020sey,Bardeen:1972fi,Tavlayan:2020cso,Cieslik:2023qdc,Bakun:2024dwq,Fujita:2009bp}. A detailed study of bound geodesics is presented in \cite{Wilkins:1972rs}. The astrophysical relevant trajectories have been studied in \cite{Rana:2019bsn}, and the null geodesics of its exterior in \cite{Gralla:2019ceu}. The radial geodesics have been also classified \cite{Compere:2021bkk} and the trajectories of massive particles are presented in \cite{Fayos:2007ks}, were a geometrical locus of all the orbits in the space of physical parameters in Kerr space-time is determined. Here, we are going to show that the condition \eqref{cond:4} is satisfied by the geodesics of the Kerr metric, and therefore a condition for the massive particle surfaces/ photon spheres  is going to be derived. We also bring a simple explanation for the Carter constant inside our formalism.\\ 
The Kerr metric in Boyer-Lindquist coordinates is written
\begin{eqnarray}\label{Kerr:1}
ds^2&=-\left(1-\frac{2Mr}{\Sigma}\right)dt^2-\frac{4 M a r\sin^2{\theta}}{\Sigma}dtd\phi\nonumber\\
&+\Sigma\left(\frac{dr^2}{\Delta}+d\theta^2\right)+\frac{\Upsilon}{\Sigma}\sin^2{\theta}d\phi^2
\end{eqnarray}

where 

\begin{eqnarray}
\Sigma &=& r^2+a^2 \cos^2(\theta),\\ 
\Delta &=& r^2-2Mr+a^2,\\
\Upsilon &=& (r^2+a^2) \Sigma +2 Ma^2 r\sin^2(\theta).
\end{eqnarray}

Replacing \eqref{Kerr:1} in \eqref{jacobim:3} we obtain
\begin{equation}\label{mKerr}
J_{ij}dx^{i}dx^{j}=F(r,\theta)\left(\frac{\Sigma}{\Delta}dr^2+\Sigma d\theta^2\right),
\end{equation}

where
\begin{equation}\label{FKerr:1}
F(r,\theta)=\delta +\frac{1}{\Delta\Sigma}\left(\Upsilon E^2+4aMr EL-\left(\csc^2(\theta)\Delta-a^2\right)L^2\right).
\end{equation}

Before applying the condition \eqref{cond:4} to the Kerr metric we calculate:

\begin{equation}\label{fS:1}
\begin{split}
F(r,\theta)\Sigma&=\delta\Sigma +\frac{\Upsilon E^2+4aMr EL}{\Delta}-\left(\csc^2(\theta)-\frac{a^2}{\Delta} \right)L^2\\
&=\delta(r^2+a^2\cos^2(\theta))+\left(\frac{(r^2+a^2)^2}{\Delta}-a^2\sin^2(\theta)\right)E^2+\frac{4MrEL}{\Delta}-\left(\csc^2(\theta)-\frac{a^2}{\Delta} \right)L^2\\
&=\delta r^2+\frac{(r^2+a^2)^2}{\Delta}E^2+\frac{4MrEL}{\Delta}+\frac{a^2}{\Delta}L^2+\delta a^2\cos^{2}(\theta)-a^2\sin^2(\theta)E^2-\frac{\cos^2{\theta}L^2}{\sin^2(\theta)}\\
&=\frac{\tilde{R}(r)}{\Delta}+\tilde{\Theta}(\theta)
\end{split}
\end{equation}

where
\begin{eqnarray}\label{RTa:1}
\tilde{R}(r)&=&\Delta(\delta r^2-(L-aE)^2)+((r^2+a^2)E-La)^2 \\
\tilde{\Theta}(\theta)&=&\delta a^2\cos^{2}(\theta)+a^2\cos^2(\theta)E^2-\frac{\cos^2(\theta)L^2}{\sin^2(\theta)},
\end{eqnarray}

Because the right side of \eqref{fS:1} has split in two functions depending on different variables and  the condition \eqref{cond:4}  $F\Sigma$ has to be a constant, thereby each function  has to be equal to the same constant $\mathcal{C}$
\begin{equation}
\frac{\tilde{R}(r)}{\Delta}=\mathcal{C}= \tilde{\Theta}(\theta),
\end{equation}
hence
\begin{equation}
\tilde{R}(r)=\mathcal{C}\Delta,\,\,\,\,\,\,\tilde{\Theta}(\theta)=\mathcal{C}.
\end{equation}

From the previous equation we find a radial function $R(r)$ and an angular function $\Theta(\theta)$ that depend on the Carter constant $\mathcal{C}$:

\begin{eqnarray}
R(r)&=&(E^2+\delta)r^4+2\delta M r^3+(E^2a^2-L^2+a^2\delta-\mathcal{C})r^2\label{RT:1}\nonumber\\
& &+2M(\mathcal{C}+(L-aE)^2)r-\mathcal{C}a^2,\\
\Theta(\theta)&=&\mathcal{C}+\cos^2(\theta)(\delta+E^2)a^2-\cot^2(\theta)L^2.\label{RT:2}
\end{eqnarray}

The $R(r)$ function is a well known result \cite{Teo:2020sey,Teo:2003ltt}, and we have obtained it by imposing that the geodesic curvature of the Riemannian metric \eqref{FKerr:1} vanishes\cite{Arganaraz:2019fup}, in other words by assuming that $F\Sigma=cte$.  Moreover, we can enforce the condition $\partial_{r}(F\Sigma)=0$, then the condition \eqref{cond:4} now transforms to 
\begin{equation}\label{condkerr:1}
R'(r)\Delta-R(r)\Delta'=0.
\end{equation} 

When there are not horizons $\Delta \neq 0$ , the condition \eqref{condkerr:1} is satisfied when 
\begin{equation}\label{rcond}
R(r)=0,\,\,\, R'(r)=0.
\end{equation}

Solving equations \eqref{rcond} using \eqref{RT:1} we obtain for circular orbits ($\theta=\pi/2$)
\begin{eqnarray}
\frac{E^2}{m^2}&=&\frac{\left(aM^{1/2}-r^{1/2}(r-2M)\right)^2}{\pm 2a M^{1/2}r^{3/2}+r^2(r-3M)},\label{kerrc:1}\\
\frac{L^2}{m^2}&=&\frac{m^2 M(a^2+2a M^{1/2}r^{1/2}+r^2)^2}{\pm 2aM^{1/2}r^{3/2}+r^2(r-3M)}.\label{kerrc:2}
\end{eqnarray}
We have written the previous expression in such a way that is easy to see that when $a=0$ the Schwarzschild case is recovered, compare with the expressions in \cite{Bardeen:1972fi}. A massive particle surface will exist if the right hand side of \eqref{kerrc:1} is a constant. Moreover, the radius of photon spheres can be deduced by setting the denominator of \eqref{kerrc:1} to zero, namely
\begin{equation}\label{kerrp:1}
\pm 2a M^{1/2}r^{3/2}+r^2(r-3M)=0.
\end{equation}
The solutions of \eqref{kerrp:1} are given by
\begin{equation}
r_{ph}=2M\left(1+\cos\left(\frac{2}{3}\textit{arcos}\left(\mp \frac{a}{M}\right)\right)\right).
\end{equation}
The innermost stable circular orbit can be found when $\frac{dE}{dr}=0$, thus when calculated for \eqref{kerrc:1} the radius of the ISCO can be found analytically. The result is a long and not particularly illuminating 
expression that we do not write. When $a=0$ we recover the radius of the ISCO for the Schwarzschild black hole $r=6M$. Similarly, When $a=0$ the photon orbit radius of the Schwarzschild spacetime is recovered. For $a=M$, the extremal case, we have $r_{ph}=M$ and $r_{ph}=4M$. In the extremal case there is not an ISCO, unless the mass $M$ becomes negative.\\

\subsubsection{Curvature and asymptotic limits}
\begin{figure}[h!]
    \centering
    \includegraphics[scale=0.55]{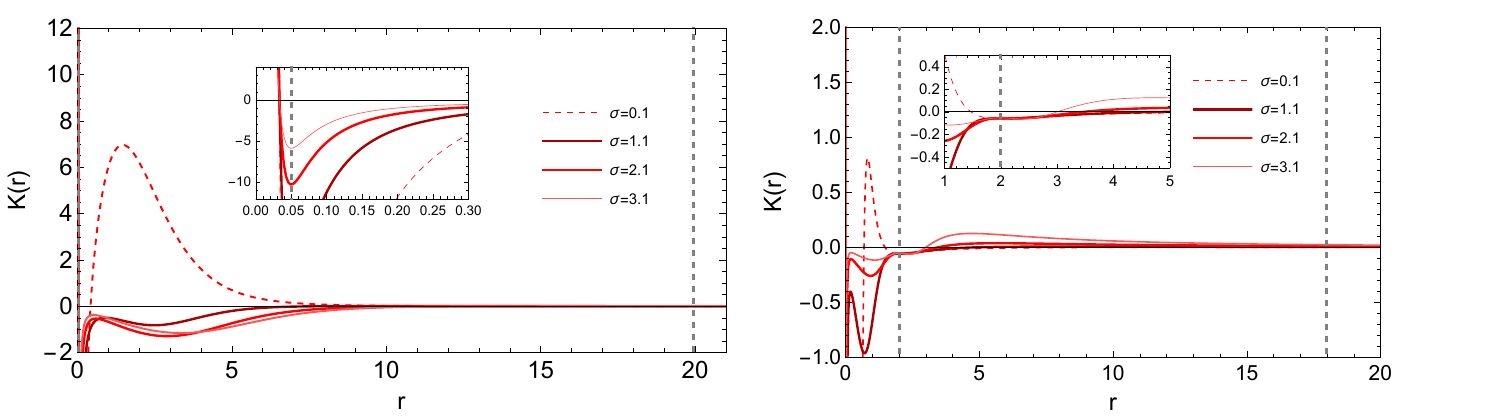}
    \caption{We have plotted the Gaussian curvature \eqref{gaussiank:1} for null trajectories of the Kerr spacetime \eqref{Kerr:1}. In the left panel we present the Gaussian curvature for different values of the impact parameter $\sigma=0.1,1.1,2.1,3.1$ with fixed energy $E=1$. We have taken $a=1$ and $M=10$. The vertical gray dashed lines represent the black hole horizons. Similarly, in the right panel we present the same Gaussian curvature as in the left panel but setting $a=0.1$ and $M=1$.}
    \label{fig:GaussianKerr}
\end{figure}

The asymptotic limits of the curvatures could give a clue about the existence of timelike circular orbits\footnote{The existence of light rings could be inferred form the same expressions by setting $\delta=0$.} TCO's. Replacing \eqref{mKerr} together with \eqref{FKerr:1} into \eqref{gaussiank:1} we get a general expression for the Gaussian curvature of the Jacobi metric of the Kerr spactime. We avoid to write the long expression but we present a plot for null trajectories of the Kerr spacetime metric \eqref{Kerr:1} in Fig. \ref{fig:GaussianKerr}. The curves are the equatorial sections ($\theta=\pi/2$) of the full Gaussian curvature. As expected,  curves corresponding to different values of the impact parameter $\sigma=0.1,1.1,2.1,3.1$  with fixed energy $E=1$ and two pairs of corresponding values of $a$ and $M$. What is evident is the behavior far away from the horizon, the Gaussian curvature tends to zero at infinity. Depending on the values of $sigma$, the Gaussian curvature will reach zero from above or from below, but in any case all the fluctuations are before we reach the exterior horizon represented by vertical gray dashed lines. Indeed, from the analytical expression of the Gaussian curvature (not shown here) we obtain:.

\begin{equation}
\lim_{r\rightarrow \infty}\mathcal{K}\approx \frac{1}{r^2}\approx 0,
\end{equation}
therefore we have an asymptotically flat Riemannian manifold. Although in Fig. \eqref{fig:GaussianKerr} we have particularized for fixed values of $a$ and $M$, the general result is independent of the parameter $a$, and therefore it is also satisfied by the Gaussian curvature of the Schwarzschild metric. On the other side, at the horizon  $r_{H}=M\pm \sqrt{M^2-a^2}$ the Gaussian curvature behaves as

\begin{eqnarray}\label{GklimitK}
\lim_{r\rightarrow r_{H}}\mathcal{K}&=& - \frac{(\Delta'(r))^2}{4 (a L (a L + 4 E M r) + E^2 \Upsilon(r,\theta))}\nonumber\\
&=&\frac{a^2-M^2}{\left(2(M^2\pm M\sqrt{M^2-a^2})E+a L\right)^2}
\end{eqnarray}

When $a=0$ we obtain $\lim_{r\rightarrow r_{H}}\mathcal{K}=-1/(16 E^2 M^2)$, which is the same expression obtained in  \cite{Bermudez-Cardenas:2024bfi} near the horizon of the Schwarzschild Jacobi metric. The Gaussian curvature in \eqref{GklimitK} is negative, it shows that we reach zero from negative values then it should grow until reaching zero.\\

The geodesic curvature behaves differently. Replacing \eqref{mKerr} and \eqref{FKerr:1} into \eqref{geode} a general expression for the geodesic curvature of the Jacobi metric coming from Kerr spacetime  is obtained, the difficult expression is not written but we have plotted the equatorial ($\theta=\pi/2$) sections of it in Fig. \ref{fig:geodesicKerr}. Note how the geodesic curvature tends to zero for big values of the radial coordinate $r$. It is important to note that when the geodesic curvature vanishes we will have a light ring. The geodesic curvature grows from negative values towards zero. The geodesic curvature does not vanishes at the horizons, and we can calculate the limit:

\begin{figure}[h]
    \centering
  \includegraphics[scale=0.65]{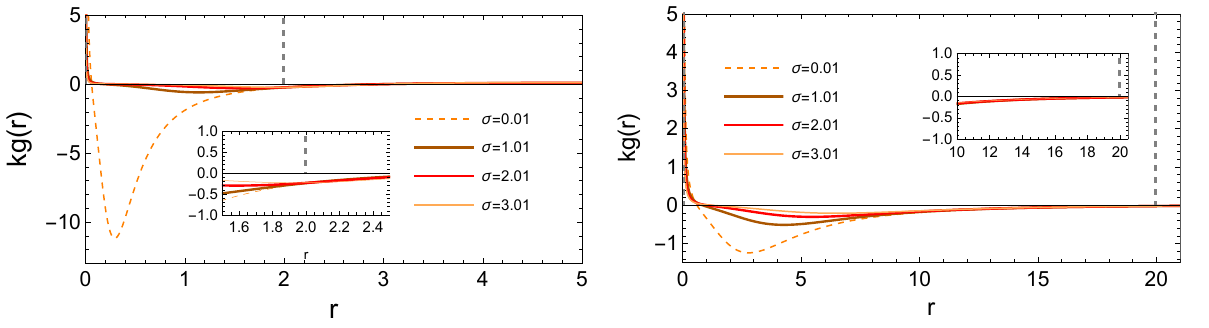}
    \caption{We have plotted the geodesic curvature \eqref{geo:1} for null trajectories of the Kerr spacetime \eqref{Kerr:1}. In the left panel we have set $a=0.1$ and $M=1$  for different values of the impact parameter $\sigma=0.1,1.1,2.1,3.1$ setting $E=1$. In the right panel we have a similar plot with $a=1$ and $M=10$. The vertical gray dashed lines represent the location of the horizons of the black hole.}
    \label{fig:geodesicKerr}
\end{figure}

\begin{eqnarray}\label{gclimitK}
\lim_{r\rightarrow r_{H}}\kappa_g&=&-\frac{\Delta'(r)}{2\sqrt{a L(a L+ 4 M E r)+E^2 \Upsilon(r,\theta)}}\\
&=&-\frac {\sqrt{M^2-a^2}}{2\left(2(M^2\pm M\sqrt{M^2-a^2})E+a L\right)}.
\end{eqnarray}
When $a=0$ we recover the Schwarzschild limit $\lim_{r\rightarrow r_{H}}\kappa_g=-1/(4ME)$ \cite{Bermudez-Cardenas:2024bfi}.  The geodesic curvature goes from a constant negative value at the horizon to zero at spatial infinity.

\subsubsection{Black hole shadows}
Black hole shadows are very important for experimental measurements. These shadows can be also studied using curvatures of the Jacobi metric. The condition for the existence of circular null/timelike orbits comes from the vanishing of the geodesic curvature. The shadows of Kerr black holes are very well known \cite{Young:1976zz, Wei:2019pjf, 1973blho.conf..215B,Perlick:2021aok}. We are going to show that they can be derived using our formalism. By setting 
\begin{equation}\label{coord:1}
\xi= \frac{L}{E},\,\,\,\,\eta=\frac{\mathcal{C}}{E^2},
\end{equation}

we are able to find a relationship between the coordinates

\begin{equation}\label{shd:0}
(a^2-a\xi+r^2)((a^2-a)M+(a^2+a)\xi-3Mr^2+r^3)=\frac{m^2}{E^2}r(a^2+r(r-2M))^2,
\end{equation}

then when $m=0$ the solution of the previous equations leads to the result \cite{Teo:2003ltt}
\begin{equation}\label{shd:1}
\xi=\frac{3Mr^2-r^3-a^2(M+r)}{a(r-M)}.
\end{equation}
Similarly, we obtain 
\begin{eqnarray}
\eta=\frac{\xi^2(r-M)+2a \xi M-2r^3-a^2(M+r)}{M-r}\nonumber\\
+\frac{m^2}{E^2}\left(\frac{a^2+r(2r-3M)}{M-r}\right).
\end{eqnarray}
Setting $m=0$, and using \eqref{shd:1}, from the previous equation we recover \cite{Teo:2003ltt}
\begin{equation}\label{shd:2}
\eta=\frac{r^3(4a^2M-r(3M-r)^2)}{a^2(r-M)^2}.
\end{equation}

Using \eqref{shd:1} and \eqref{shd:2} celestial coordinates for an observer, at the inclination angle $\sigma$, can be defined
\begin{eqnarray}
\alpha&=&-\xi \csc(\sigma),\label{coord:2}\\
\beta&=&\pm\sqrt{\eta+a^2\cos^2(\sigma)-\xi^2\cot^2(\sigma)}.\label{coord:3}
\end{eqnarray}
The coordinates $\alpha,\beta$ constitute a parametrization of the form of the shadows, a detailed study can be found in \cite{Wei:2019pjf,Wei:2018xks,Tsukamoto:2014tja,Tsukamoto:2017fxq}. We have shown that the information regarding the geodesic of Kerr spacetime is encoded in the Riemannian metric \eqref{jacobim:3}. 

\subsection{Kerr-de Sitter spacetime}
We extend the analysis to a metric that is axi-symmetric and stationary but it is asymptotically (A)de Sitter. It is a natural extension of the Kerr metric but now with an extra term that is responsible of asymptotic change. In Boyer-Lindquist coordinates the Kerr-(A)dS  metric is written \cite{AlBalushi:2019obu}

\begin{eqnarray}\label{kerrds:1}
ds^2&=&-\frac{1}{\Sigma^2}\left(\Delta_r-a^2\Delta_{\theta}\sin^{2}(\theta)\right)dt^2-\frac{2a}{\Sigma^2 \Gamma}\sin^2(\theta)(\Delta_{\theta}(r^2+a^2)-\Delta_r)dt d\phi\\
& &+\frac{\Sigma^2}{\Delta_r}dr^2+\frac{\Sigma^2}{\Delta_{\theta}}d\theta^2
+\frac{\sin^2(\theta)}{\Sigma^2\Gamma^2}\left((r^2+a^2)^2\Delta_{\theta}-a^2\Delta_r\sin^2(\theta)\right)d\phi^2
\end{eqnarray}

where
\begin{eqnarray}
\Delta_r&=&(r^2+a^2)\left(1+\frac{r^2}{\epsilon \ell^2}\right)-2Mr,\\
\Delta_{\theta}&=&1-\frac{a^2\cos^2(\theta)}{\epsilon \ell^2},\\
\Gamma&=&1-\frac{a^2}{\epsilon \ell^2},\\
\Sigma^2&=&r^2+a^2\cos^2(\theta).
\end{eqnarray}
with $\epsilon=+1$ for AdS and $\epsilon=-1$ for dS spacetimes. When $\ell\rightarrow \infty $ we recover the Kerr metric \eqref{Kerr:1}. Replacing \eqref{kerrds:1} in expression \eqref{jacobim:3} we obtain the Jacobi metric

\begin{equation}\label{jmKdS:1}
J_{ij}dx^{i}dx^{j}=F(r,\theta)\left(\frac{\Sigma^2}{\Delta_r}dr^2+\frac{\Sigma^2}{\Delta_{\theta}}d\theta^2\right)
\end{equation}
where

\begin{eqnarray}
F(r,\theta)=\delta+\left(\frac{1}{\Sigma^2}\left(\frac{(r^2+a^2)^2}{\Delta_r}-\frac{a^2\sin^2(\theta)}{\Delta_{\theta}}\right)E^2
\right.\nonumber\\
\left.
+\frac{2a \Gamma}{\Sigma^2}\frac{r^2+a^2}{\Delta_r}-\frac{1}{\Delta_{\theta}}\right)EL\nonumber\\
+\frac{\Gamma^2}{\Sigma^2}\left(\frac{a^2}{\Delta_r}
-\frac{1}{\Delta_\theta \sin^2(\theta)}\right)L^2
\end{eqnarray}

In order to use the condition \eqref{cond:4} we calculate
\begin{equation}
F\Sigma^2=\tilde{R}(r)+\tilde{\Theta}(\theta),
\end{equation}
where
\begin{eqnarray}\label{rtKads:1}
\tilde{R}(r)&=&\delta r^2+\frac{1}{\Delta_r}\left((r^2+a^2)E^2+\Gamma aL\right)^2\\
\tilde{\Theta}(\theta)&=&\delta a^2 \cos^2(\theta)-\frac{1}{\Delta_{\theta}}\left(a\sin^2(\theta)E+\frac{\Gamma L}{\sin^2(\theta)}\right)^2.
\end{eqnarray}
The condition \eqref{cond:4} tell us that the product $F\Sigma^2$ is constant and therefore, each of the functions $\tilde{R}(r)$ and $\tilde{\Theta}(\theta)$ have to be equated to the same constant $\mathcal{C}_2$, then we define
\begin{eqnarray}\label{rtKads:2}
R_2(r)&=&-\Delta_r\mathcal{C}_2+\Delta_r\delta r^2+\left((r^2+a^2)E^2+\Gamma aL\right)^2\\
\Theta_2(\theta)&=&-\Delta_\theta\mathcal{C}_2+\Delta_\theta \delta a^2 \cos^2(\theta)\nonumber\\
& &-\left(a\sin^2(\theta)E+\frac{\Gamma L}{\sin(\theta)}\right)^2.
\end{eqnarray}
Moreover, getting the derivative of $F(r)\Sigma^2$ we arrive at the equations that need to be satisfied by the circular orbits, the condition analogous to \eqref{rcond}, it reads
\begin{equation}
R_2(r)\Delta'_{r}+R_2'(r)\Delta_r=0,
\end{equation}
which implies that the system $R_2(r)=0,R'_2(r)=0$ is solved for circular timelike geodesic. Solving the system we arrive to
\begin{eqnarray}\label{em:1}
\frac{E^2}{m^2}&=&\frac{1}{r^2 \Delta_r}\left(\frac{\Delta_r}{2}+\frac{r\Delta'_r}{4}\right)^2,\\
\frac{L^2}{m^2}&=&\frac{1}{\Gamma^2 a^2}\left(\frac{m^2}{E^2}\left(\frac{\Delta_r}{2}+\frac{r\Delta'_r}{4}\right)^2+\frac{E^2}{m^2}(r^2+a^2)^2
\right.\nonumber\\
& &
\left.
+2(r^2+a^2)\left(\frac{\Delta_r}{2}+\frac{r\Delta'_r}{4}\right)\right),
\end{eqnarray}
then, if the right side of \eqref{em:1} is constant  there is a massive particle surface. The previous expressions can be written explicitly:

\begin{eqnarray}
\frac{E^2}{m^2}&=&\frac{\left(2a^2r^2+3r^4+(a^2-2Mr+r^2)\epsilon \ell^2\right)^2}{4r^4(a^2+r^2)\epsilon\ell^2+4r^2(a^2-2Mr+r^2)\epsilon \ell^4}, \label{emkads:1}\\
\frac{L^2}{m^2}&=&\frac{1}{\ell^2 a^2}\frac{\left(r^2(r^2+a^2)(2a^2+5r)+\epsilon \ell^2\left(a^4+r^3(4r-7M)+a^2r(5r-3M)\right)\right)^2}{4\epsilon \ell^2 r^2(r^4+\epsilon \ell^2 r(r-2M)+a^2(r^2+\epsilon \ell^2))}.\label{emkads:2}
\end{eqnarray}

If the denominator of the expression \eqref{emkads:1} is set to zero we find the condition for the existence of a photon sphere:
\begin{equation}
r^4+(a^2+\epsilon \ell^2)r^2-2\epsilon \ell r+a^2\epsilon \ell^2=0.
\end{equation}
The previous equation is in accordance with the Fig. \ref{fig:GaussianKerrdSAds}. Depending on the values of $\sigma$ the Gaussian curvature will diverge at $\pm \infty$. In any case, there is no way that the Gaussian curvature tends to zero at infinity. This is a consequence of the non-asymptotically flat nature of the Kerr-(A)dS spacetime.

\subsubsection{Curvature and asymptotic limits}
\begin{figure}[h]
    \centering
 \includegraphics[scale=0.55]{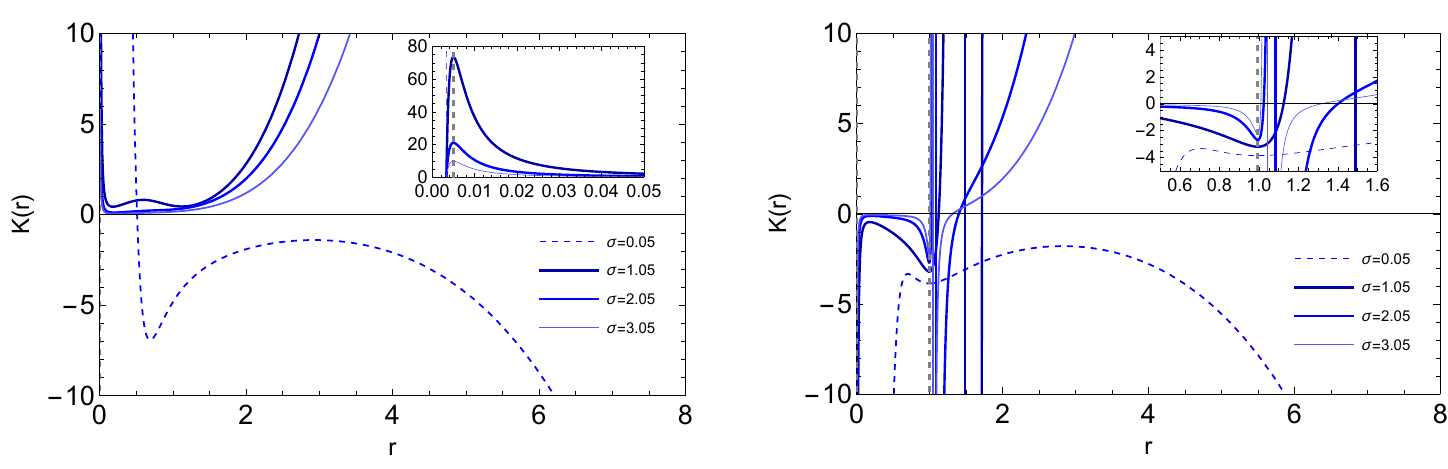}
    \caption{The graphic represents the Gaussian curvature corresponding to the null trajectories of the Kerr-Ads/dS spacetime metric \eqref{kerrds:1}. In the left panel we present the Gaussian curvature of the Kerr-Ads ($e=-1$), for different values of the impact parameter $\sigma=0.05,1.05,2.05,3.05$ with fixed energy $E=1$. We have set $a=0.1$ and $M=1$. In the right panel we present the same Gaussian curvature corresponding to the null geodesics with $a=1$ and $M=10$. The vertical gray dashed lines represent the horizons of the black hole. }
    \label{fig:GaussianKerrdSAds}
\end{figure}

The Gaussian and geodesic curvatures of \eqref{jmKdS:1} can be calculated. As before, the asymptotic limits will provide a hint of the geometric behavior of the metric. In the asymptotic limit the Gaussian of \eqref{jmKdS:1} behaves as:
\begin{equation}
\lim_{r \rightarrow \infty}\mathcal{K}=\frac{-(1+a^2l^2)e^2l^2\delta+L^2 \left(a^2-e l^2\right)^2-a^2e^2l^4}{\delta ^2 e^3 l^6}
\end{equation} 
The $2$-dimensional surface is no longer flat at infinity, this is a direct consequence of the asymptotic limits of the Kerr-(A)dS metric. In Fig. \ref{fig:GaussianKerrdSAds} we have plotted the Gaussian curvature of the null trajectories of the Kerr-(A) dS spacetime metric. In the left panel the Gaussian curvature corresponds to Kerr-AdS spacetime. We can see that now the asymptotic limit is not easily defined as in the asymptotically flat spacetimes. Depending on the values of the impact parameter $\sigma$ the Gaussian curvature will change its asymptotic limit. Physically, it implies that as far as we go from the horizon the null orbits become more unstable/stable. Moreover, there are not circular null/timelike orbits after the horizon. In the right panel we can see the behavior of the Gaussian curvature of the de Sitter Jacobi metric. Near the horizon the behavior is a little bit chaotic, but far from it we have a similar behavior to that of Kerr-anti de Sitter, namely Gaussian curvature going to very high absolute values. In both cases the degree of stability/instability of null trajectories will be increased.

\begin{figure}[h]
    \centering
 \includegraphics[scale=0.65]{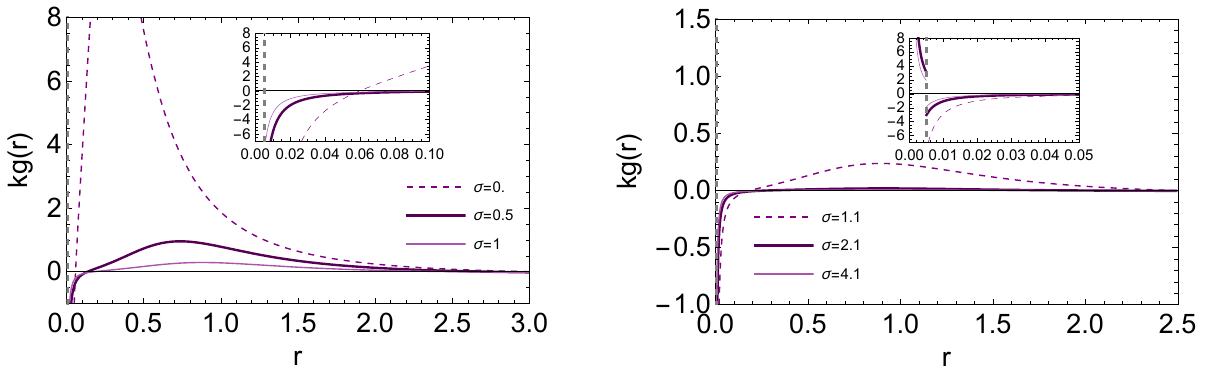}
    \caption{We plot the geodesic curvature for the null trajectories of the Kerr-Ads spacetime metric \eqref{kerrds:1}. We have set for both panels $a=0.1$ and $M=1$, and the impact factor $\sigma=0,0.5,1,1.1,2.1,4.1$. The dependence on the value of $\sigma$ is evident. For all curves we take $E=1$.}
    \label{fig:geoKerrads}
\end{figure}

The geodesic curvature of Kerr-AdS Jacobi metric for null trajectories is plotted in Fig. \ref{fig:geoKerrads}. The behavior at infinity is similar to the case of Kerr spacetime which is asymptotically flat spacetime, However, the interesting things are happening before. In the left panel it can be seen that there are points where the geodesic curvature vanishes, leading to the existence of light rings. In the right panel of Fig. \ref{fig:geoKerrads} we can find points where the geodesic curvature also vanishes. Note that the value of $\sigma$ is important, for values $\sigma<1$ it is clear the vanishing of the geodesic curvature.

\begin{figure}[h]
    \centering
 \includegraphics[scale=0.65]{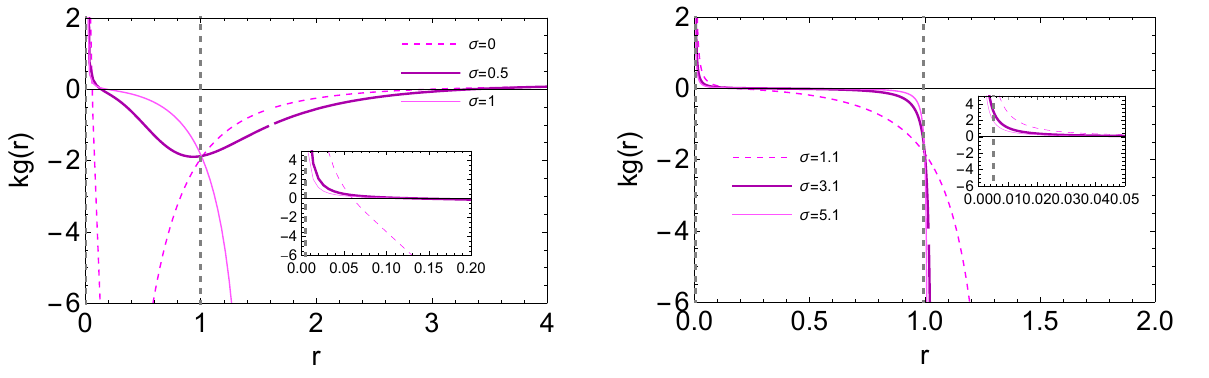}
    \caption{We plot the Gaussian curvature for the null trajectories of the Kerr-de Sitter spacetime \eqref{kerrds:1}. In both panels we have set $a=0.1$ and $M=1$. the values of the impact parameter $\sigma=0,0.5,1,1.1,3.1,5.1.$ The vertical gray dashed lines are located to represent the horizons of the metric. The value of the energy $E=1$ is fixed.}
    \label{fig:geoKerrds}
\end{figure}

The behavior of Kerr-dS is different for $\sigma>1$. In the right panel of Fig. \ref{fig:geoKerrds} the geodesic curvature in the dS case is plotted for values of $\sigma>1$, it is clear that in this case the geodesic curvature never reaches zero after the horizon. In the left panel the cases $\sigma<1$ are plotted. Here, contrary tho the opposite case, the geodesic curvature reaches zero beyond the horizon. Therefore, the value of $\sigma$ is pivotal for the existence of light rings

\begin{figure}[h]
    \centering
 \includegraphics[scale=0.45]{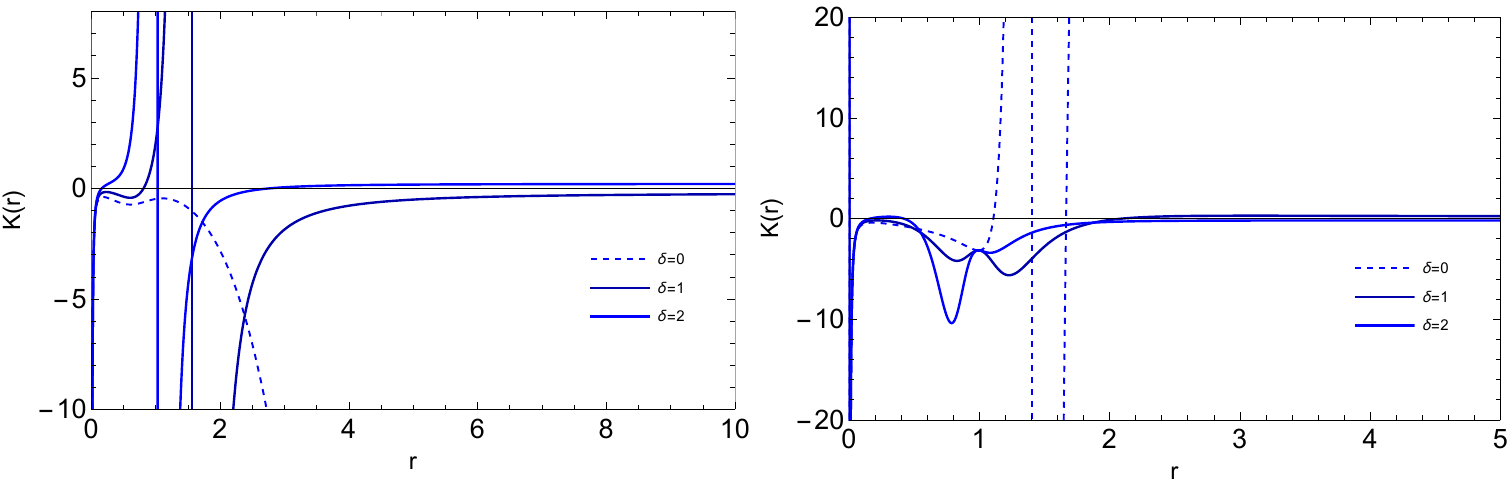}
    \caption{We plot the Gaussian curvature for the timelike trajectories of the Kerr-AdS and Kerr-dS spacetimes \eqref{kerrds:1}. In the left panel the corresponding Gaussian curvature for Kerr-AdS. The timelike cases are depicted in blue non-dashed curves ($\delta=1,\delta=2$) and the null case correspond to the dashed curve ($\delta=0$). In the right panel we plot the Gaussian curvature for the Kerr-dS case. We have set $a=0.1$ , $M=1$, $\sigma=0.1$,$l=1$ and $E=1$ in both plots.}
    \label{fig:Kadsdsmassive}
\end{figure}

\begin{figure}[h]
    \centering
 \includegraphics[scale=0.45]{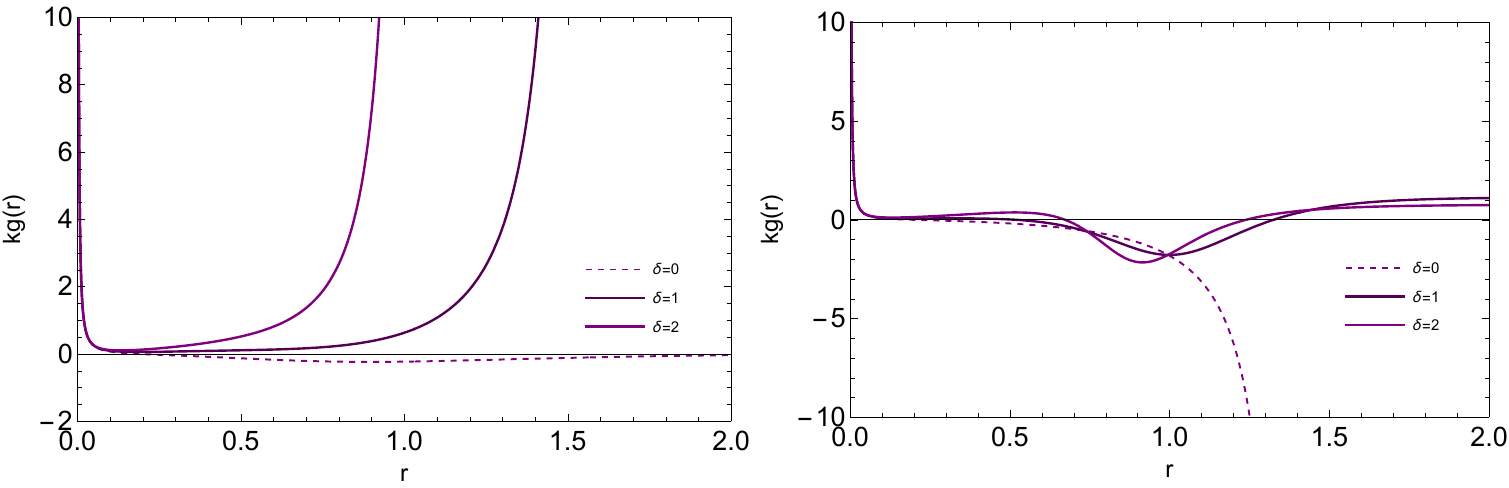}
    \caption{We plot the geodesic curvature for the timelike trajectories of the Kerr-AdS and Kerr-dS spacetimes \eqref{kerrds:1}. In the left panel the corresponding Gaussian curvature for Kerr-AdS. The timelike cases are depicted in purple non-dashed curves ($\delta=1,\delta=2$) and the null case correspond to the dashed curve ($\delta=0$). In the right panel we plot the geodesic curvature for the Kerr-dS case. We have set $a=0.1$ , $M=1$, $\sigma=0.1$,$l=1$ and $E=1$ in both plots.}
    \label{fig:kgadsdsmassive}
\end{figure}

The power of our method relies in its simplicity but also in the geometry of surfaces. In all previous figures we have shown how the Gaussian and the geodesic curvature for null trajectories behave. Contrary to what happens with the traditional approach, generalization to timelike trajectories is straightforward. In Fig. \ref{fig:Kadsdsmassive}  we have plotted the Gaussian curvature for the timelike trajectories of the Kerr-AdS and Kerr-dS spacetimes \eqref{fig:kgadsdsmassive}, in left and right panel respectively. The timelike cases are depicted using continuous curves ($\delta=1,2$) and the null case ($\delta=0$) is depicted using a dashed curve. Note that the presence of extra terms corresponding to the massive case changes the asymptotics of both curvatures. For $\delta\neq 0$ the asymptotic limit of the Gaussian curvature is not necessarily $\pm\infty$, instead it converges to a constant value, meaning that at some point after the horizon the timelike orbits do not change its stability. Similarly, the geodesic curvature changes its behavior for timelike orbits. In the left panel of Fig. \ref{fig:kgadsdsmassive} the existence of light rings an be inferred form the zeros of the curve, on the other side there is no timelike circular orbits. In the right panel, there is not light rings, but there are timelike circular orbits. We are able to determine the existence of null/timelike circular orbits without resting in the solution of the geodesic equation, even the stability of orbits can be studied only by geometric methods.

In the near horizon limit we have
\begin{equation}
\lim_{r \rightarrow r_{H}}\mathcal{K}=\left.\frac{\left(4r^3+2(a+\epsilon \ell^2)r-2\epsilon \ell M\right)^2}{2 \epsilon^2 \ell^4\left((r^2+a^2)E+\Gamma a L\right)^2 } \right\vert_{r=r_{h}}.
\end{equation} 
The Gaussian curvature has to grow from negative values towards positive ones for timelike geodesics, therefore there should be at least a point where it vanishes.\\
The geodesic curvature at infinity behaves as 
\begin{equation}
\lim_{r\rightarrow\infty}\kappa_{g}=\frac{1}{l (\delta e)^{1/2}},
\end{equation}
which shows that at infinity there is not timelike circular orbits. However, note that in the null case $\delta=0$ the geodesic curvature vanishes at infinity, meaning that there is a null circular orbit at infinity. On the other side near the horizon we get
\begin{equation}
\lim_{r \rightarrow r_{H}}\kappa_g=\left.\frac{\left(4r^3+2(a+\epsilon \ell^2)r-2\epsilon \ell M\right)}{2 \epsilon \ell^2\left((r^2+a^2)E+\Gamma a L\right) } \right\vert_{r=r_{h}}.
\end{equation}
Now it is easy to see that for big values of $M$ the previous equation is going to be negative and since at infinity it is positive there should be at least one point where it vanishes, leading to the existence of TCO's a very similar behavior to that of the Kerr metric.

\subsubsection{Black hole shadows}
As in the Kerr case, the shadows of Kerr-(A)dS are known, however we are going to derive them from the geodesic curvature of circular geodesics, which comes from $\partial_r F\Sigma^2=0$. This condition translates to $R_2(r)=0,R'_2(r)=0$, therefore for $L/E$ and $C/E^2$  we should find the expression for the parameters that define the shadow, thus we obtain
\begin{eqnarray}
\frac{L}{E}&=&\frac{m^2}{E^2}\frac{1}{a\Gamma}\left(\frac{\Delta_r}{2}+\frac{r\Delta'_r}{4}\right)-\frac{r^2+a^2}{\Gamma a}\\
\frac{\mathcal{C}}{E^2}&=&4\Gamma a r\frac{L}{E}+4r(r^2+a^2)-\frac{m^2}{E^2}
\left(2\frac{\Delta_r}{\Delta'_r}+r^2\right)
\end{eqnarray}

The previous expression can be written explicitly, and the shadows of the black holes can be studied using the coordinates $\alpha$ y $\beta$ defined in \eqref{coord:2} and \eqref{coord:3} respectively.\\
The results are in accordance with the literature, but the most important message about this derivation y that the geodesic curvature of a two dimensional metric encodes the information regarding the shadows. We do not need to solve the geodesic equations.

\section{Discussion}

We have extended the geometric approach for massive particle surfaces presented in \cite{Bermudez-Cardenas:2024bfi}. We have studied the massive particle surfaces using a Riemannian metric: the Jacobi metric. We have deduced the master equation that provides a criteria for the existence of massive particle surfaces for metrics of the type  \eqref{staticm:1} and for the stationary case where the metric is given by \eqref{staticm:2}. The condition for the existence of circular orbits, which in our formalism leads to the master equation whose constancy leads to the existence of a massive particle surface, was obtained from the vanishing of the geodesic curvature of the Jacobi metric. The condition \eqref{cond:4} is an important result, it is the generalization of the umbilicity property for stationary spacetimes. The calculations are more involved compared with the static asymptotically flat case studied in \cite{Bermudez-Cardenas:2024bfi}. The Jacobi metric for stationary spacetimes is of the Randers-Finsler type and in order to obtain a Riemannian metric we have used the approach developed in \cite{Arganaraz:2021fwu}. We have also shown how to calculate the radius of the photon sphere which in some cases needs to be calculated numerically. The examples that we have chosen are prototypical for stationary spacetimes with different asymptotics. We have also shown that all results reduce to the null surfaces when $m=0$. The only ingredients needed are the conformal factor of the Jacobi metric and the component $g_{\theta\theta}$ of the spacetime metric. Our formalism was able to recover the shadows of the proposed black hole metric. We have shown that the asymptotic limit does not represent an inconvenient for our results, we have applied the formalism to asymptotically (A)dS. We have calculated the expressions for $\xi=L/E$ and $\eta=\mathcal{C}/E^2$ that allows us to study the shadows of the black hole. Our Jacobi metric, apart form encoding the information about geodesics, it stores the information regarding the massive particle surfaces/ photon surfaces and the shadows of the black hole.  Our approach is completely geometric, and all results come from intrinsic curvatures of the Jacobi metric. An interesting problem would be to study different geometries, such as wormholes \cite{Arganaraz:2019fup,Duenas-Vidal:2022kcx}. Moreover, the Jacobi metric  provides a tool for studying the existence of TCO's. Being a purely geometric approach it opens the door to use the Riemannian geometry for characterizing all physical properties of a spacetime. In Appendix \ref{app} we study a more involved metric, a family of rotating black holes which are solutions of $4-$dimensional Einstein-Maxwell-dilaton (EMD) gravity. The results are similar to the previous case, however the calculations are lengthy. 

\appendix
\section{Appendix}\label{app}
\subsection{Einstein-Maxwell-dilaton metric}

As we have seen in the case of Kerr-(A)dS spacetime, our formalism can be applied to a stationary spacetimes which are not asymptotically flat. Here we apply the results of section \ref{s:4} to a family of rotating black holes which are solutions of $4-$dimensional Einstein-Maxwell-dilaton (EMD) gravity \cite{Sheykhi:2007bv,Ghosh:2007jb,Li:2024oyc}. The metric reads:

\begin{eqnarray}\label{wsol:1}
ds^2=-W(r)dt^2+\frac{dr^2}{W(r)}-2a f(r)\sin^2(\theta)dtd\phi+r^2 R^2(r)(d\theta^2+\sin^2(\theta)d\phi^2)
\end{eqnarray}

where

\begin{eqnarray}
W(r)&=&-M r^{2\gamma-1}+\frac{(1+\alpha^2)}{(1-\alpha^2)}\left(\frac{r}{b}\right)^{2\gamma}\nonumber\\
& &-\frac{\Lambda (1+\alpha^2)^2}{(2(3-\alpha^2))}\frac{r^{2(1-\gamma)}}{b^{-2\gamma}}+q^2(1+\alpha^2)\frac{r^{2\gamma-2}}{b^{2\gamma}},\\
f(r)&=&M b^{2\gamma}r^{2\gamma-1}+\frac{\Lambda}{2}\frac{(1+\alpha^2)^2}{(3-\alpha^2)}\frac{r^{2(1-\gamma)}}{b^{-2\gamma}}\\
& &-q^2 (1+\alpha^2)\frac{r^{2\gamma-2}}{b^{2\gamma}},\\
\Phi(r)&=&\frac{\alpha}{(1+\alpha^2)}\ln\left(\frac{b}{r}\right),\\
R(r)&=&e^{\alpha \Phi},\\
h(r)&=&r^{-1}.
\end{eqnarray}

with  $M$ being the mass of the black hole, $q$ its charge and $a$ its angular momentum. The metric \eqref{wsol:1} is a solution of EMD theory if the component of the electromagnetic potential $A_{\phi}$ and the potential of the dilation field $V(\Phi)$ are given by
\begin{eqnarray}
A_{\phi}= a q h(r)\sin^{2}(\theta)\\
V(\Phi)=\Lambda_{o} e^{\lambda_o}+\Lambda e^{\lambda \Phi},
\end{eqnarray}  
with
\begin{equation}
\Lambda_o=\frac{2\alpha^2}{b^2(\alpha^2-1)},\,\,\,\,\,\,\lambda_o=\frac{2}{\alpha},\,\,\,\,\,\,\lambda= 2\alpha.
\end{equation}

Replacing \eqref{wsol:1} in the Jacobi metric \eqref{jacobim:3} we get
\begin{equation}
J_{ij}dx^{i}dx^{j}=F(r,\theta)\left(\frac{dr^2}{W(r)}+r^2 R^2(r)d\theta^2\right),
\end{equation}
where

\begin{equation}\label{F:2}
F(r,\theta)=\left(\delta+\frac{2 a ELf(r)\sin^2(\theta)-E^2 r^2 R^2(r)\sin^2(\theta)+L^2 W(r)}{-4a^2f^2(r)\sin^4(\theta)-r^2 R^2\sin^2(\theta)W(r)}\right).
\end{equation}

Although the calculation can be done in full generality, we focus in the slow rotating case \cite{Ghosh:2007jb}, namely $a\neq 0, a^2\approx 0$, then the conformal factor \eqref{F:2} becomes

\begin{equation}
F(r,\pi/2)_{a^2\approx 0}=\left(\delta+\frac{r^2 R^2(r)E^2-2a EL f(r)-L^2\csc^2(\theta) W(r)}{r^2 R^2(r)W(r)}\right).
\end{equation}

In order to apply the condition \eqref{cond:4}, we need to calculate $F(r,\theta)g_{\theta\theta}$ first, thus
\begin{eqnarray}\label{gF:1}
g_{\theta\theta}F(r,\pi/2)_{a^2\approx 0}&=&L^2\csc^2(\theta)-\delta r^2 R(r)^2\nonumber\\
& &+\frac{2a EL f(r)}{W(r)}-\frac{E^2 r^2 R^2(r)}{W(r)}\nonumber\\
&=&\tilde{\Theta}(\theta)+\tilde{R}(r).
\end{eqnarray}
where 
\begin{eqnarray}
\tilde{\Theta}(\theta)&=&L^2\csc^{2}(\theta),\\
\tilde{R}(r)&=&-\delta r^2 R(r)^2+\frac{2a EL f(r)}{W(r)}-\frac{E^2 r^2 R^2(r)}{W(r)}.
\end{eqnarray}

The condition $g_{\theta\theta}F(r,\pi/2)_{a^2\approx 0}=cte$ leads us to define a new constant $\mathcal{C}_3$ such that
\begin{eqnarray}\label{tr:1}
\Theta_3(\theta)&=&\tilde{\Theta}(\theta)-\mathcal{C}_3,\\
R_3(r)&=&\tilde{R}(r)-\mathcal{C}_3.
\end{eqnarray}
The constant $\mathcal{C}_3$ is called the Carter constant. By applying the condition \eqref{cond:4} it is direct to find the condition for the existence of massive particle surfaces. However, although  $E^2/m^2$ can be calculated explicitly the result is a long and complicated expression, but the existence of massive particle surfaces is guaranteed. Moreover, the black hole shadows can be calculated, indeed by setting $\theta=\pi/2$ from \eqref{tr:1} we obtain $\mathcal{C}_3=L^2$. Using this expression in $g_{\theta\theta}F(r,\pi/2)_{a^2\approx 0}=0$ and in $\partial_{r}F(r,\theta)=0$ with \eqref{le:1} we get

\begin{equation}\label{le:1}
\frac{L}{E}=\frac{r R(r)(2W(r)(1+\delta W(r))(R(r)+r R'(r))-rR(r)W'(r))}{2a (W(r)f'(r)-f(r)W'(r))}
\end{equation}
\begin{equation}\label{ce:1}
\frac{K}{E}=\frac{r R(r)(r R(r)(1+\delta W(r))f'(r)-f(r)(2r R'(r)+R(r)(2+\delta r W'(r))))}{W(r)f'(r)-f(r)W(r)}
\end{equation}

The expressions \eqref{le:1} and \eqref{ce:1} are the generalization to massive case of the equations (28) in \cite{Li:2024oyc}. Using the coordinates defined in \eqref{coord:1} we can build the shadows of the black hole using \eqref{coord:2} and \eqref{coord:3}.

\textbf{Acknowledgements}

The work of BBC was funded by the National Agency for Research and Development (ANID)/ Scholarship Doctorado Nacional 2022/ folio 21220518.

\printbibliography

\end{document}